\documentclass[adp,fleqn]{w-art}
\usepackage{times}
\usepackage{w-thm}
\usepackage[]{graphicx}
\usepackage{amssymb}
\begin{document}
\DOIsuffix{theDOIsuffix}
\Receiveddate{}
\Reviseddate{}
\Accepteddate{}
\Dateposted{}




\title[Quantum Gravity]{Quantum Gravity: General Introduction and Recent
Developments}

\author[C. Kiefer]{Claus Kiefer
  \footnote{E-mail:~\textsf{kiefer@thp.uni-koeln.de}, 
            Phone: +49\,221\,470\,4301, 
            Fax: +49\,221\,470\,2189}}
\address{Institut f\"ur Theoretische Physik, Universit\"at zu
K\"oln, Z\"ulpicher Strasse 77, 50937 K\"oln, Germany}


\begin{abstract}
I briefly review the current status of quantum gravity.
After giving some general motivations for the need of such a theory,
I discuss the main approaches in quantizing general relativity:
Covariant approaches (perturbation theory, effective theory,
and path integrals) and canonical approaches (quantum geometrodynamics,
loop quantum gravity). I then address quantum gravitational aspects
of string theory. This is followed by a discussion of black holes and
quantum cosmology. I end with some remarks on the observational status
of quantum gravity.
\end{abstract}


\maketitle                   

\tableofcontents 


\section{Why quantum gravity?}

The consistent implementation of the gravitational interaction into
the quantum framework is considered to be {\em the} outstanding
problem in fundamental physics. A big obstacle so far is the
lack of a direct experimental hint.
Therefore, the motivations for the need of such a theory are 
at present purely
theoretical. These motivations strongly indicate, however, that
the present framework of theoretical physics is incomplete. 
In the following I shall briefly discuss the main reasons that
motivate the search for a quantum theory of gravity. More details on this
as well as on the other issues discussed in this review can be found
in my monograph \cite{OUP} to which I refer the reader for more details
and references.

Why should one believe that there is a need for a quantum theory of gravity?
\begin{itemize}
\item {\em Singularity theorems}: Under very general conditions, it follows
from general relativity (GR) that
singularities in spacetime are unavoidable \cite{HP}. GR,
therefore, predicts its own breakdown. This applies in particular to the
universe as a whole: there are strong indications for the presence of an
initial singularity (this can be inferred from the existence
of the cosmic microwave background radiation).
 Since the classical theory is then no longer
applicable, a more comprehensive theory must be found -- the general
expectation is that this is a {\em quantum} theory of gravity.
\item {\em Initial conditions in cosmology}: This is related to the 
first point. Cosmology as such is incomplete if its beginning cannot
be described in physical terms. It seems that even an inflationary epoch
cannot avoid a singularity in the past \cite{BGV}. Inflation itself is
based on the `no-hair conjecture' according to which spacetime approaches
locally de~Sitter space if a positive (effective) cosmological constant
is present. An implicit assumption for the validity of the
no-hair conjecture is that modes smaller than the 
Planck scale (see below) are not excited to macroscopic scales
(`trans-Planckian problem').
\item {\em Evolution of black holes}: Black holes radiate with a temperature
proportional to $\hbar$, the Hawking temperature \cite{HP}, see below.
For the final evaporation, a full theory of quantum gravity is needed
since the semiclassical approximation leading to the Hawking temperature
then breaks down. 
\item {\em Unification of all interactions}: All non-gravitational interactions
have so far been successfully accomodated into the quantum framework.
Gravity couples universally to all forms of energy. One would therefore expect
that in a unified theory of all interactions, also gravity is described
in quantum terms.
\item {\em Inconsistency of an exact semiclassical theory}:
All attempts to construct a fundamental theory where a classical gravitational
field is coupled to quantum fields have failed 
up to now \cite{OUP}. Such a 
semiclassical theory seems to exist only in an approximate sense.
\item {\em Avoidance of divergences}: It has long been speculated that
quantum gravity may lead to a theory devoid of the ubiquitous divergences
arising in quantum field theory. This may happen, for example, through
the emergence of a natural cutoff at small distances (large momenta).
In fact, modern approaches such as string theory or loop quantum gravity
(see below) provide indications for a discrete structure at small scales.
\end{itemize}
At which scale(s) would one expect that effects of quantum gravity
necessarily occur? As shown by Max Planck in 1899, gravitational constant $G$,
quantum of action $\hbar$ and speed of light $c$ can be combined
in a unique way (apart from numerical factors) to provide fundamental
units of length, time, and mass, respectively.
These `Planck units' read
\begin{eqnarray}
l_{\rm P} &=& \sqrt{\frac{\hbar G}{c^3}} \approx 1.62 \times 10^{-33} 
                                         \,{\rm cm}\ , \\
t_{\rm P} &=& \sqrt{\frac{\hbar G}{c^5}} \approx 5.40 \times10^{-44} 
                                         \,{\rm s}\ , \\
m_{\rm P} &=& \sqrt{\frac{\hbar c}{G}} \approx 2.17\times 10^{-5} \, {\rm g}
                                       \approx 1.22 \times 10^{19} \, 
                                         {\rm GeV} \ .
\end{eqnarray}
One might think that the presence of a fundamental length scale is not
a Lorentz invariant notion. This is, however, not necessarily the case because
special relativity cannot be applied in this regime and, moreover, lengths 
should be related to eigenvalues of a geometric quantum operator, cf.
\cite{length} for a detailed discussion in the framework of loop 
quantum gravity. 

\begin{figure}[h]
\label{fig_berlin05_1}
\caption{Different levels for the application of quantization}
\begin{center}
\includegraphics[width=4cm]{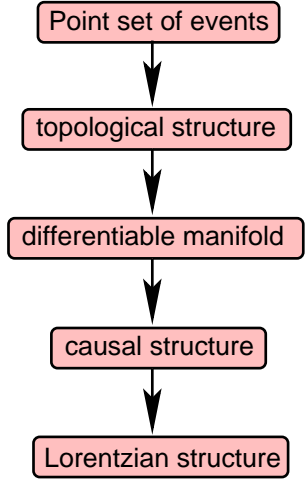}
\end{center}
\end{figure}

Quantization of gravity means quantization of geometry.
But which structures should be quantized, that is, to which structures
should one apply the superposition principle? Chris Isham has
presented the hierarchy of structures depicted in Fig.~\ref{fig_berlin05_1}
\cite{CI94}. 
Structures that are not quantized remain as absolute (non-dynamical)
entities in the formalism. One would expect that in a fundamental theory
no absolute structure remains. This is referred to as {\em background
independence} of the theory.
\newpage
\bigskip
\centerline{\bf Problem of time}
\bigskip

A particular aspect of background independence is the `problem of time',
which arises in any approach to quantum gravity. On the one hand,
time is external in ordinary quantum theory; the parameter $t$
in the Schr\"odinger equation,
\begin{equation}
{\rm i} \hbar \, \frac{\partial \psi}{\partial t} =  H \psi\ , 
\end{equation}
is identical to Newton's absolute time---it is {\em not} turned into an
operator and is presumed to be prescribed
from the outside. This is true also in
special relativity where the absolute time
$t$ is replaced by Minkowski spacetime,
which is again an absolute structure.

On the other hand,
time in general relativity is dynamical because it is part of spacetime
described by Einstein's equations,
\begin{equation}
R_{\mu\nu} - \frac{1}{2} \, g_{\mu\nu} \, R 
 +\Lambda g_{\mu\nu}= \frac{8\pi G}{c^4} 
T_{\mu\nu}\ .
\end{equation}
Both concepts cannot be true fundamentally. It is expected that
the fundamental theory does not contain any absolute structure.
We shall see below how far present approaches to quantum gravity
implement background independence.
 
\bigskip
\centerline{\bf Hawking effect}
\bigskip

The relationship between quantum theory and gravity 
has been experimentally investigated only on the level of
Newtonian gravity implemented into the Schr\"odinger equation
or its relativistic generalizations, cf. \cite{KW} for a recent review.
On the level of quantum field theory on a classical gravitational
background there exists a specific prediction
which has, however, not yet been tested: Black holes
radiate with a \emph{temperature} 
proportional to $\hbar$ (`Hawking radiation'),
\begin{equation}
\label{TBH1}
T_{\rm BH}=\frac{\hbar\kappa}{2\pi k_{\rm B}c}\ ,
\end{equation}
where $\kappa$ is the surface gravity. 
For a Schwarzschild black hole this yields
\begin{equation}
\label{TBH}
T_{\rm BH} =\frac{\hbar c^3}{8\pi k_{\rm B}GM} 
\approx 6.17\times 10^{-8}
 \left(\frac{M_{\odot}}{M}\right)\ {\rm K}\ .
\end{equation}
The black hole shrinks due to Hawking radiation and possesses a finite
lifetime. The final phase, where $\gamma$-radiation is being emitted,
could be observable.
The temperature \eqref{TBH}
is unobservably small for black holes that result
from stellar collapse. One would need primordial black holes produced in
the early universe because they could possess a 
sufficiently low mass \cite{Carr}. For example,
black holes with an initial mass of $5\times 10^{14}$ g would evaporate
at the present age of the universe. In spite of several attempts, no
experimental hint for black-hole evaporation has been found. 
Primordial black holes can result from density fluctuations produced
during an inflationary epoch \cite{Carr}. However, they can only be
produced in sufficient numbers if the scale invariance of the power
spectrum is broken at some scale, cf. \cite{BBKP}.

Since black holes radiate thermally, they also possess an \emph{entropy},
the `Bekenstein--Hawking entropy', which is given by the expression
\begin{equation}
\label{SBH}
S_{\rm BH}=\frac{k_{\rm B}c^3A}{4G\hbar}=k_{\rm B}\frac{A}{4l_{\rm P}^2}
 \ ,
\end{equation}
where $A$ is the surface area of the event horizon.
For a Schwarzschild black hole with mass $M$, this reads
\begin{equation}
S_{\rm BH}\approx 1.07\times 10^{77}k_{\rm B}\left(\frac{M}{M_{\odot}}\right)^2
\ .
\end{equation}
Since the Sun has an entropy of about $10^{57}k_{\rm B}$, this means 
that a black hole resulting from the collapse of a star with a few solar
masses would experience an increase in entropy by twenty orders of
magnitude during its collapse.

It is one of the challenges of any approach to provide a microscopic
explanation for this entropy, that is, to derive \eqref{SBH} from a counting
of microscopic quantum gravitational states.

\bigskip
\centerline{\bf Main Approaches}
\bigskip

What are the main approaches?
\begin{itemize}
\item {\em Quantum general relativity}: The most straightforward
attempt, both conceptually and historically, is the application
of `quantization rules' to classical general relativity. One distinguishes
\begin{itemize}
 \item {\em Covariant approaches}: These are approaches that employ
 four-dimensional covariance at some stage of the formalism.
 Examples include perturbation theory, effective field theories,
 re\-nor\-ma\-li\-za\-tion-group approaches, and path integral methods.
 \item {\em Canonical approaches}: Here one makes use of a Hamiltonian
formalism and identifies appropriate canonical variables and conjugate
momenta. Examples include quantum geometrodynamics and loop quantum gravity.
\end{itemize}
\item {\em String theory}: This is the main approach to construct
a unifying quantum framework of all interactions. The quantum aspect of
the gravitational field only emerges in a certain limit in which the
different interactions can be distinguished.
\item There are a couple of other attempts such as 
  quantization of topology, or the theory of causal sets, which I 
will not address in this short review.
\end{itemize}

\section{Covariant quantization}

\bigskip
\centerline{\bf Perturbative Quantum Gravity}
\bigskip

Historically, the first attempt to quantize gravity was through 
perturbation theory. In 1930, L\'eon Rosenfeld investigated 
the gravitational field produced by an electromagnetic field and
calculated the gravitational self-energy after quantization \cite{Rosenfeld}.
It turned out to be infinite. Infinities have been encountered before
in the self-energy of the electron, and Rosenfeld did his calculation
(following a question by Werner Heisenberg) to see whether such infinities
already occur in situations where only `fields' 
(no `matter fields' such as electrons) are present.

Perturbation theory starts by decomposing the metric, $g_{\mu\nu}$, into a
background part, $\bar{g}_{\mu\nu}$, and a `small' perturbation,
$f_{\mu\nu}$,\footnote{We set $c=1$ from now on.}
\begin{equation}
\label{Feynman}
g_{\mu\nu}=\bar{g}_{\mu\nu}+\sqrt{32\pi G}f_{\mu\nu}\ .
\end{equation}
The important assumption is the presence of an (approximate)
\emph{background} with respect to which standard perturbation theory
(formulation of Feynman rules, etc.) can be applied.
In this approximate framework the quantum aspects of gravity
are encoded in a spin-2 particle propagating on the background --
the {\em graviton}.
In contrast to Yang--Mills theory, however, the ensuing perturbation
theory is {\em non-renormalizable}: At each order in the expansion
with respect to $G$, new types of divergences occur which have to be
absorbed into appropriate parameters
that have to be fixed by measurement. All together, one would have to
introduce infinitely many parameter, rendering the theory useless.
An explicit calculation of the 
\emph{2--loop divergence} in 1985 has shown that the divergences
are real and that there is no `miraculous' cancellation of divergences
already in the absence of non-gravitational fields
\cite{GS}.\footnote{Pure gravity is finite at one-loop order.} 
The corresponding divergent two-loop Lagrangian reads
\begin{equation}
{\mathcal L}^{({\rm div})}_{\rm 2-loop}=
\frac{209\hbar^2}{2880}\,\frac{32\pi G}{(16\pi^2)^2\epsilon} \,
\bar{R}^{\alpha\beta}{}_{\gamma\delta}\,
\bar{R}^{\gamma\delta}{}_{\mu\nu}\,
\bar{R}^{\mu\nu}{}_{\alpha\beta}\ ,
\end{equation}
where $\epsilon$ is a regularization parameter (from dimensional
regularization) that goes to zero and thus produces a divergence.
The presence of supersymmetry (SUSY) alleviates the occurrence of
divergences; they, however, appear at higher loops and thus do not seem to
prevent the theory from being perturbatively non-renormalizable.  

\bigskip
\centerline{\bf Effective Field Theories and Renormalization Group}
\bigskip

Even if the theory is perturbatively non-renormalizable,
it may lead to \emph{definite predictions} at low energies. 
This is the framework of `effective field theories'.
As an example I quote the
quantum gravitational correction to the Newtonian potential,
calculated in \cite{BDH},
\begin{equation}
V(r)=-\frac{Gm_1m_2}{r} \, \left(1+3\frac{G(m_1+m_2)}{2r}
+\frac{41}{10\pi} \, \frac{G\hbar}{r^2}\right)\ .
\end{equation}
This result is independent of the ambiguities that are present
at higher energies. In fact, effective field theories are successfully
applied elsewhere, for example, 
`chiral perturbation theory' in QCD
(where one considers the limit of
the pion mass $m_{\pi}\to 0$) is such an effective theory.

A theory that is perturbatively non-renormalizable may still be
non-perturbatively renormalizable. A standard example is the
three-dimensional Gross--Neveu model with a large number of
flavour components \cite{CVMS}. Could this happen for
quantum general relativity, too? 
A theory may be `asymptotically safe' in the sense that
a non-Gaussian (non-vanishing) 
ultraviolet (UV) fixed point exists non-perturbatively
\cite{Weinberg}. In fact, there are indications that this is the
case \cite{LR}: Consider an `Einstein--Hilbert truncation' 
defined by the action (in $d$ spacetime dimensions)
\begin{equation}
\Gamma_{k}[g] =
\frac{1}{16\pi G_{k}} \, \int {\rm d}^d x \, \sqrt{g} \,
\left(-R(g)+2\bar \lambda_{k}\right) \,,
\end{equation}
plus classical gauge fixing, where $\bar \lambda_{k}$ and $G_{k}$ are the
momentum ($k$-) dependent cosmological and gravitational constant,
respectively. Studying the renormalization-group equations within this
truncation (which is of course a very restrictive ansatz), it is found
in \cite{LR} that the dimensionless versions of these parameters,
$\lambda_k=k^{-2}\bar \lambda_{k}$ and $g_k=k^{d-2}G_k$, have an
UV attractive non-Gaussian fixed point at values $(\lambda_*,g_*)\neq (0,0)$.
Therefore, as $k\to \infty$,
\begin{equation}
G_{k} \approx \frac{g_*}{k^{d-2}} \,,
\quad
\bar\lambda\equiv\Lambda\approx \lambda_* \, k^2\ .
\end{equation}
The gravitational constant thus vanishes in this limit 
(for $d>2$), and the theory would
be asymptotically free. Using more general truncations, it may also
be possible to obtain a 
small positive cosmological constant as a strong infrared quantum effect
\cite{RS} and to get a growth of $G$ at large distances which could
explain the observed flat galaxy rotation curves \cite{RW}.
Although these results have been obtained within special truncations
and may not survive in the full theory, they show that quantum GR
could in principle
be perturbatively renormalizable and lead to testable
predictions even at a macroscopic level (and what could be more
macroscopic than galaxies). 
This is also the hope for the
approaches to quantum GR presented below.

\bigskip
\centerline{\bf Path Integrals}
\bigskip

In quantum mechanics and quantum field theory, path integrals 
provide a convenient tool for a wide range of applications. Among them 
are saddle point approximations and non-perturbative approaches,
the latter typically in a lattice framework.
 In quantum gravity, a path-integral formulation would have to
employ a sum over all four-metrics for a given topology,
\begin{equation}
\label{Zg}
Z[g]=\int{\mathcal D}g_{\mu\nu}(x)\, 
{\rm e}^{{\rm i} S[g_{\mu\nu}(x)]/\hbar}\ .
\end{equation}
In addition, one would expect that a sum over all topologies 
has to be performed. Since four-manifolds are not classifiable, this
is an impossible task. Attention is therefore restricted to a given
topology (or to a sum over few topologies). Still, the evaluation
of an expression such as \eqref{Zg} meets great mathematical and
conceptual difficulties. 

 From a methodological point of view,
one distinguishes between a Euclidean and a Lorentzian version of the
path integral \eqref{Zg}. In the Euclidean version, one performs a
Wick rotation $t\to -{\rm i}\tau$ in the action and integrates over
Euclidean metrics only. In ordinary quantum field theory, this improves
the convergence properties of the path integral. In quantum gravity, however,
the situation is different because the Euclidean action is unbounded from
below. This unboundedness is not necessarily a problem as long as
one is only concerned with a saddle point approximation of the path integral,
that is, a WKB aproximation of the form $\exp(-I)$, where $I$ denotes
the extremal value of the Euclidean action. 

\begin{figure}[h]
\label{fig_berlin05_2}
\caption{Hartle--Hawking instanton: The dominating contribution to the
Euclidean path integral is assumed to be half of a four-sphere attached
to a part of de Sitter space.}
\begin{center}
\includegraphics[width=7cm]{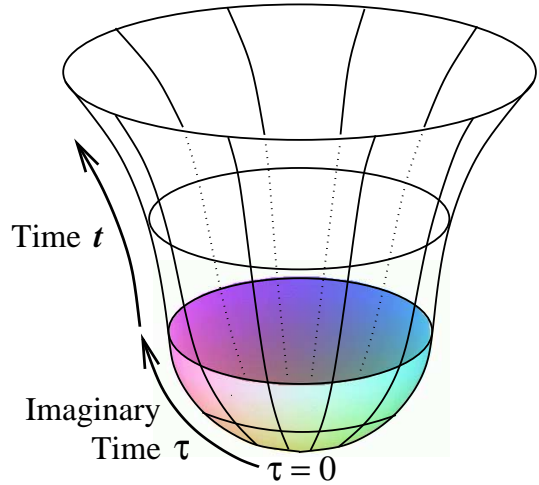}
\end{center}
\end{figure}

Such an approximation lies, in fact,
at the heart of one of the prominent proposals for boundary conditions
in quantum cosmology---the `no-boundary condition' or `Hartle--Hawking
proposal' \cite{HH}. This consists of two parts. First, it is assumed
that the Euclidean form of the path integral is fundamental, and that the
Lorentzian structure of the world only emerges 
in situations where the saddle point is
complex. Second, it is assumed that one integrates over metrics with
one boundary only (the boundary corresponding to the present universe),
so that no `initial' boundary is present; this is the origin of the
term `no-boundary proposal'. The absence of an initial boundary
is implemented through appropriate regularity conditions.   
In the simplest situation, one finds as the dominating geometry 
the `Hartle--Hawking instanton' depicted in Fig.~\ref{fig_berlin05_2}:
the dominating contribution at small radii is (half of the) Euclidean
four-sphere $S^4$, whereas for bigger radii it is (part of) de Sitter space,
which is the analytic continuation of $S^4$. Both geometries are matched
at a three-geometry with vanishing extrinsic curvature.
The Lorentzian nature of our universe would thus only be an `emergent'
phenomenon. 

In more general situations, one has to look for integration contours
in the space of complex metrics that render the integral convergent.
Consider one quantum cosmological example discussed in
detail in \cite{Annals}. It describes a scalar field 
(called $\chi$ after a field redefinition) conformally coupled
to the scale factor, $a$, of a Friedmann universe. The classical solutions
are confined to a rectangle centred around the origin in the
$(a,\chi)$-plane. One can explicitly construct wave packets that follow
these classical trajectories. (They are solutions of the 
Wheeler--DeWitt equation discussed below.)
The corresponding quantum states are
normalizable in both the $a$ and the $\chi$ direction. A general 
quantum gravitational (cosmological) path integral depends on two values
for $a$ and $\chi$, called $a',a''$ and $\chi',\chi''$.
respectively. 
Fig.~3 shows the corresponding $(a'',\chi'')$ space;
the values for $a'$ and $\chi'$ define the origin of the `light cones'
depicted in the Figure by bullets.

The Hartle--Hawking wave function is obtained for $a'=0=\chi'$, that is, 
for the case when the
light cones in Fig.~3 shrink to one cone 
centred at the origin. For this model, the path integral can be
evaluated {\em exactly}.
The investigation in \cite{Annals} has shown that
there is no contour for the path integration
in the complex-metric plane that leads to a wave function which can be used
in the construction of wave packets following the classical trajectories:
the resulting wave functions either diverge along the `light cones'
or they diverge for large values of $a$ and $\chi$. The states are
thus not normalizable, and it is not clear how they should be
interpreted.

\begin{figure}[h]
\label{fig_berlin05_3}
\caption{The wave functions obtained from the
path integral for the model investigated in
\cite{Annals} either diverge along the light cones 
in minisuperspace or for large 
values of $a$ and $\chi$. They exhibit oscillatory behaviour in the
shaded regions.}
\begin{center}
\includegraphics[width=8cm]{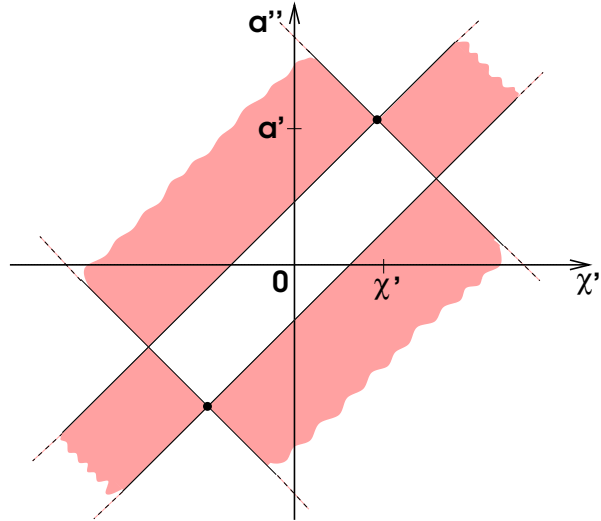}
\end{center}
\end{figure}

\bigskip
\centerline{\bf Dynamical triangulation}
\bigskip

An alternative method to attack the full path integral \eqref{Zg}
is dynamical triangulation, cf.
\cite{AJL} and the references therein. Here one starts from a 
\emph{Lorentzian} path integral in the first place and employs a
discretization of the geometry
 (a Euclidean formulation is used in an intermediate step
of the calculations). In contrast to earlier attempts such
as Regge calculus, the 
edge lengths of simplices are held {\em fixed}; the sum is then
performed over all possible
manifold gluings of equilateral simplices,
using Monte Carlo simulations.

Some interesting results have already been obtained. 
First, the Hausdorff dimension $H$ of space, defined by
$V(r)\propto \langle r\rangle^H$, where $V$ is the space volume,
is found to read
 $H=3.10\pm 0.15$. This is an indication for the
three-dimensionality of space (and thus for the four-dimensionality
of spacetime). Since there is no background in quantum gravity,
the dimension $d+1$ of the geometries to be summed over in the path integral
does not need to be identical to the resulting value for $H+1$. 
Strangely enough, one had found earlier that for a Euclidean path integral
the result is always $H=2$ for $d>2$. Second, a positive value for
the cosmological constant, 
$\Lambda >0$, is needed. This is in accordance with results
from recent measurements of the cosmic acceleration. However, no
special value for $\Lambda$ seems to be preferred.
Third, for large scales the volume seems to behave semiclassically.
However, in spite of these promising hints, a continuum limit is still elusive.

\section{Canonical quantization}

One of the earliest attempts to quantize gravity employs a
Hamiltonian approach. In a first step, the classical theory
(GR) is reformulated in `3+1 form' in which it
describes the dynamics of three-geometries (with matter fields on them).
Global hyperbolicity of the spacetime from which one starts is a
necessary prerequisite. It must be emphasized that the topology
of the three-dimensional space $\Sigma$ is fixed, that is, there is one
canonical theory for each topology.
In a second step, the canonical variables, that is, the configuration
variable and its momentum (the `symplectic structure') is chosen.
The third step then addresses the canonical quantization procedure.
The most straightforward approach would be to choose the three-metric
and its conjugate momentum (which is a linear function of the
second fundamental form). The resulting quantum theory is called
`quantum geometrodynamics' and is historically the oldest approach.
More recently, alternative systems of
variables have proved to be of great interest;
because they can be interpreted as a connection or a loop variable,
the corresponding approaches are called `quantum connection dynamics'
or `loop quantum gravity'. I shall review these approaches briefly
in the following.   

The central equations are the \emph{quantum constraints}. The
invariance of the classical theory under coordinate transformation leads
to four (local) constraints: the Hamiltonian constraint,
\begin{equation}
\label{WDW}
{\mathcal H}_{\bot}\Psi=0\ ,
\end{equation}
and the three diffeomorphism (or momentum) constraints,
\begin{equation}
\label{momentum}
{\mathcal H}_a\Psi=0\ .
\end{equation}
The total gravitational Hamiltonian reads (apart from boundary terms)
\begin{equation}
H=\int {\rm d}^3x\ \left(N{\mathcal H}_{\bot}+N^a{\mathcal H}_a\right) \ ,
\end{equation}
where $N$ (`lapse function') and $N^a$ (`shift vector') are
Lagrange multipliers.
In the connection and loop approaches, three additional (local) constraints
emerge because of the freedom to choose the local triads upon which the
formulation is based. 

Some interesting features occur for 2+1-dimensional spacetimes
\cite{Carlip} but I shall restrict myself in the following to the
four-dimensional case.

\bigskip
\centerline{\bf Quantum geometrodynamics}
\bigskip

The canonical variables are the three-dimensional metric, $h_{ab}(x)$,
and its conjugate momentum.\footnote{In addition, there are of course
additional non-gravitational fields, which I shall not mention 
explicitly in the following.}
 In this case one usually refers to
\eqref{WDW} (or the full equation $H\Psi=0$)
as the `Wheeler--DeWitt equation'. Here are some characteristics
of this approach:

\begin{itemize}
\item The wave functional  
$\Psi$ depends on the \emph{three}-dimensional metric, but 
  because of \eqref{momentum} it is invariant
      under coordinate transformations on three-space.
\item No external time parameter is present anymore -- in this
       sense the theory is `timeless'. This also holds for the
      connection and loop approaches. 
\item Such constraints result from any theory that is classically
      reparametrization invariant, that is, a theory without
      background structure.
\item The Wheeler--DeWitt equation
is (locally) hyberbolic, and one can thereby define a local `intrinsic time'
 distinguished by the sign in this wave equation.
\item This approach is a candidate for a non-perturbative quantum theory
of gravity. But even if it is superseded by a more fundamental theory
at the Planck scale (such as superstring theory, see below), it 
should approximately be valid away from the Planck scale. The reason is
that GR is then approximately valid, and the quantum theory
from which it emerges in the WKB limit is quantum geometrodynamics.
\end{itemize}

Let us consider a simple example: a
Friedmann universe with scale factor $a\equiv
  {\rm e}^{\alpha}$ containing a
massive scalar field {$\phi$}. This is an
interesting model for quantum cosmology (see below). In this case,
the diffeomorphism constraints are identically fulfilled, and the
Wheeler--DeWitt equation \eqref{WDW} reads 
\begin{equation}
\label{mini}
H\psi\equiv\left({G}\hbar^2\frac{\partial^2}
{\partial \alpha^2}
-\hbar^2\frac{\partial^2}{\partial\phi^2}
+m^2\phi^2{\rm e}^{6\alpha}-\frac{{\rm e}^{4\alpha}}{{G}}
\right)\psi(\alpha,\phi)=0\ .
\end{equation}
This equation is simple enough to find solutions (at least numerically)
and to study physical aspects such as the dynamics of wave packets
and the semiclassical limit \cite{OUP}.

The semiclassical approximation can be conveniently discussed also for
the full Wheeler--DeWitt equation, at least in a formal sense
(i.e., treating functional derivatives as if they were ordinary
derivatives and neglecting the problem of anomalies). The discussion
is also connected to the question:
Where does the imaginary unit in the 
(functional) Schr\"odinger equation come from \cite{Barbour}?
The full Wheeler--DeWitt equation is real,
\begin{equation}
\label{wdwfull}
H\Psi = 0 \ ,
\end{equation}
and one would also expect real solutions for $\Psi$.
An 
approximate solution is found through a Born--Oppenheimer-type of
scheme, in analogy to molecular physics. This is discussed at great length
in \cite{OUP}. The state then assumes the form
\begin{equation}
\label{expiS}
\Psi \approx \exp({\rm i}S_0[h]/\hbar) \, \psi[h,\phi]\ ,
\end{equation}
where $h$ is an abbreviation for the three-metric, and
$\phi$ stands for non-gravitational fields. In short, one has
\begin{itemize}
\item $S_0$ obeys the Hamilton--Jacobi equation for the gravitational field
and thereby defines a classical spacetime which is a solution to
Einstein's equations (this order is formally similar to the recovery
of geometrical optics from wave optics via the eikonal equation).
\item $\psi$ obeys an approximate (functional) Schr\"odinger equation,
\begin{equation}
\label{semi}
 {\rm i}\hbar \, \underbrace{\nabla \, S_0 \, \nabla
\psi}_{\frac{\displaystyle\partial\psi}{\displaystyle\partial t}} 
\approx H_{\rm m} \, \psi \ ,
\end{equation}
where $H_{\rm m}$ denotes the Hamiltonian for the non-gravitational fields
$\phi$. Note that the expression on the left-hand side of \eqref{semi} 
is a shorthand notation for an integral over space, in which $\nabla$
stands for functional derivatives with respect to the three-metric. 
Semiclassical time $t$ is thus defined in this limit from the
dynamical variables. 
\item The next order of the Born-Oppenheimer scheme yields
 quantum gravitational correction terms proportional to the inverse
Planck mass squared, ${1}/{m_{\rm P}^2}$. The presence of such terms
may in principle lead to observable effects, for example, in the
anisotropy spectrum of the cosmic microwave background radiation. 
\end{itemize}
The canonical formalism can also be extended to quantum supergravity
\cite{OUP,Death}.
Its semiclassical approximation -- recovery of the functional
Schr\"odinger equation and calculation of quantum gravitational correction
terms -- has recently been performed in \cite{KLM}. 

The Born--Oppenheimer expansion scheme distinguishes a state of the form
\eqref{expiS} from its complex conjugate. In fact, in a generic situation
both states will decohere from each other, that is, 
they will become dynamically independent
\cite{deco}. This is a type of symmetry breaking, in analogy to the
occurrence of parity violating states in chiral molecules. It is through
this mechanism that the i in the Schr\"odinger equation emerges.

The recovery of the Schr\"odinger equation \eqref{semi} raises an
interesting issue. It is well known that the notion of Hilbert space
is connected with the conservation of probability (unitarity) and thus
with the presence of an external time (with respect to which the
probability is conserved). The question then arises whether the concept
of a Hilbert space is still required in the {\em full} theory where
no external time is present. It could be that this concept makes sense
only on the semiclassical level where \eqref{semi} holds. 

A major problem with quantum geometrodynamics is the lack of 
a precise mathematical framework for the full theory. This provides
one with a motivation to look for alternatives. 

\bigskip
\centerline{\bf Connection and loop dynamics}
\bigskip 

Instead of the metric formulation of the last subsection one can
use a different set of variables, leading to the connection or the
loop formulation. Detailed expositions can be found, for example, in
\cite{Rovelli,Thiemann,AL}, see also \cite{OUP}. Starting point are
the `new variables' introduced by Abhay Ashtekar in 1986,
\begin{eqnarray}
E_i^a(x) &=& \sqrt{h} \, e^a_i(x)\ ,\\
GA^i_a(x) &=& \Gamma^i_a(x) + {\beta}K^i_a(x)\ .
\end{eqnarray}
Here, $e^a_i(x)$ is the local triad (with $i$ being the internal index),
$h$ is the determinant of the three-metric, $\Gamma^i_a(x)$ 
the spin connection, and $K^i_a(x)$ the second fundamental form.
The parameter $\beta\in{\mathbb C}\backslash\{0\}$ denotes a
quantization ambiguity similar to the $\theta$-parameter ambiguity in QCD
and is called the `Barbero--Immirzi parameter'. The canonical pair of
variables are the densitized triad $E_i^a(x)$ (this is the new momentum)
and the connection $A^i_a(x)$ (the new configuration variable).
They obey the Poisson-bracket relation
\begin{equation}
 \{A^i_a(x),E_j^b(y)\}=8\pi{\beta}\delta^i_j
 \delta^b_a\delta(x,y)\ . 
\end{equation}
The use of this pair leads to what is called `connection dynamics'. 
One can rewrite the above constraints in terms of these variables
and subject them to quantization.
In addition, one has to treat the `Gauss constraint' arising from the
freedom to perform arbitrary rotations in the local triads. 

The \emph{loop variables}, on the other hand, are constructed from
a non-local version of the connection. Consider a loop in the
three-space $\Sigma$, see Fig.~4.

\begin{figure}[h]
\label{fig_berlin05_4}
\caption{An oriented loop in the three-manifold $\Sigma$.}
\begin{center}
\includegraphics[width=6cm]{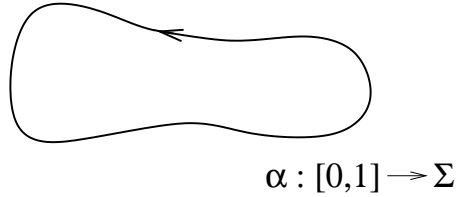}
\end{center}
\end{figure}
\noindent The fundamental loop variable is the holonomy $U[A,\alpha]$ defined
as the path-ordered product
\begin{equation}
U[A,\alpha]={\mathcal P}\exp\left(G\int_\alpha A\right)\ .
\end{equation}
The conjugate variable is the `flux' of $E^a_i$ 
through a two-dimensional surface ${\mathcal S}$ in $\Sigma$.

\bigskip
\centerline{\bf Loop quantum gravity}
\bigskip 

A central concept in the formulation of loop quantum gravity is
the {\em spin-network basis}. This is a complete orthonormal basis
with respect to which all appropriate quantum states can be expanded.
How is it defined? 
A spin network is a triple $S(\Gamma,\vec{j},\vec{N})$, where $\Gamma$ 
denotes a graph, $\vec{j}$ is a set of `spins' (representations of
SU(2)) that are attached to the curves forming the graph, and
$\vec{N}$ denotes the chosen collection of basis elements at the nodes
where the curves meet. A spin-network state is then
\begin{equation}
 \Psi_S[A]=f_S(U_1,\ldots,U_n)\ ,
\end{equation}
where $f$ is a `cylindrical function' attributed to the graph, which is
defined as a mapping from $[{\rm SU}(2)]^n$ (putting a holonomy on each curve)
to ${\mathbb C}$. One can show that a natural Hilbert-space structure 
can be constructed on the {\em kinematical} level, that is, before 
the constraints \eqref{WDW} and \eqref{momentum} are imposed. 

The introduction of the group SU(2)---and with it the algebra of
angular momenta---introduces a discrete structure into the formalism.
It is thus of interest to consider geometrical operators and discuss
their spectrum with respect to the kinematical Hilbert space. As one may
have expected, the spectrum turns out to be discrete. As one example
I want to mention the
\emph{quantization of area}. Applying the `area operator' 
$\hat{A}({\mathcal S})$ (whose classical analogue is the area of the
two-dimensional surface ${\mathcal S}$) to a spin-network state one
obtains
\begin{equation}
\hat{A}({\mathcal S})\Psi_S[A]=8\pi\beta l_{\rm P}^2
\sum_{P\in S\cap{\mathcal S}}\sqrt{j_P(j_P+1)}\Psi_S[A]\ .
\end{equation}
The points $P$ denote the intersections of the spin network with
the surface, see Fig.~5 for an example with four curves.

\begin{figure}[h]
\label{fig_berlin05_5}
\caption{A spin network $S$ intersecting the surface $\mathcal S$.}
\begin{center}
\includegraphics[width=8cm]{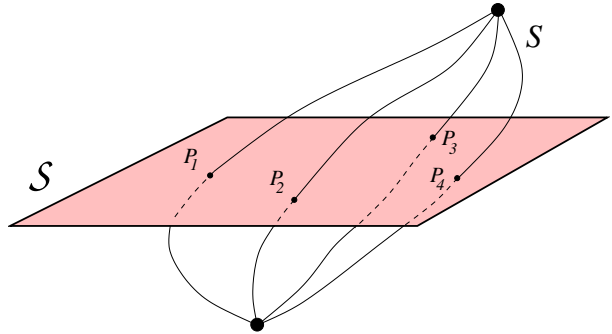}
\end{center}
\end{figure}

Note that the area operator is not invariant under three-dimensional
diffeomorphisms. 
This is because it is defined for an abstract surface
in terms of coordinates. It does also not commute with the Hamiltonian
constraint. An area operator that {\em is} invariant should be defined
intrinsically with respect to curvature invariants or matter fields.
A concrete realization of such an operator remains elusive.
  
In spite of much promising aspects, some open problems remain,
cf. \cite{HPZ} and \cite{Rovelli,Thiemann}.
They include:
\begin{itemize}
\item The presence of quantum anomalies in the constraint algebra
 could make the formalism inconsistent. How can anomalies be avoided?
\item How does one have to treat the constraints? The Gauss constraint is
easy to implement, but there are various subtleties and ambiguities with
the diffeomorphism and Hamiltonian constraints.
\item Can a semiclassical limit be obtained? Since the constraints assume
now a form different from the one in geometrodynamics, 
a Born--Oppenheimer approach cannot be straightforwardly applied.
An important role may be played by generalized coherent states. 
For large enough scales, the semiclassical approximations of 
loop quantum gravity and quantum geometrodynamics should of course coincide. 
\end{itemize}

A path-integral formulation of loop quantum gravity employs the notion
of `spin foam', that is, time-developed spin networks that are integrated
over, cf. \cite{Rovelli} for details. 

\section{String theory}

String theory (resp. M-theory) provides a drastically different approach to
quantum gravity. The idea is to first construct a quantum theory
of {\em all} interactions (a `theory of everything') from which
separate quantum effects of the gravitational field follow in some
appropriate limit. String theory transcends local field theory in that it
contains at the fundamental level
higher dimensional objects (not only strings, but also branes)
instead of points. However, as the discussion of loop quantum gravity above
shows, non-local entities can also emerge in quantum GR.
This is a consequence of background independence, which is not the
case in standard field theories such as QED.
A detailed exposition of this topic can be found in \cite{Polchinski},
cf. also \cite{Witten} for a short non-technical overview.
The main characteristics of string theory are:

\begin{itemize}
\item The appearance of gravity is inevitable. The graviton
 is an excitation of closed strings, and it appears via virtual
 closed strings in open-string amplitudes. 
\item String theory has as important ingredients the concepts of
gauge invariance, supersymmetry (SUSY), and the presence of higher dimensions.
\item It provides a unification of all interactions.
\item String perturbation theory seems to be finite at each order, 
but the sum diverges.
\item Only \emph{three} fundamental constants are present: 
 $\hbar, c$ and the string length $l_{\rm s}$; all other physical parameters
(masses and coupling constants) should in principle be derivable from them.
\item A net of dualities connects various string theories and indicates
the presence of an underlying non-perturbative theory (M-theory)
from which the various string theories can be recovered in appropriate limits.
\end{itemize}

Starting point is the formulation of a string action on the
two-dimensional worldsheet of the string. It is given by the
(generalized) Polyakov action describing the propagation of the string
in a $D$-dimensional embedding (target) spacetime.
For the bosonic string it reads
\begin{eqnarray}
S\ &\propto& l_{\rm s}^{-2}\int{\rm d}^2\sigma\
\left(\sqrt{h}h^{\alpha\beta}\partial_{\alpha}X^{\mu}\partial_{\beta}
X^{\nu}{g_{\mu\nu}(X)}\right. \nonumber \\
& & \;\;\left.
-l_{\rm s}^2\sqrt{h}\,{}^{(2)}\!R{\Phi(X)}+
\epsilon^{\alpha\beta}\partial_{\alpha}X^{\mu}\partial_{\beta}X^{\nu}
{B_{\mu\nu}(X)}\right)\ .
\end{eqnarray}
Here, $g_{\mu\nu}$ is the $D$-dimensional metric of the embedding space,
$\Phi$ is the dilaton field, and $B_{\mu\nu}$ is an 
antisymmetric tensor field. They are all background fields, that is, they
will not be integrated over in the path integral. In the path integral, 
\begin{equation}
\label{pathintegral}
Z=\int{\mathcal D}X{\mathcal D}h\ {\rm e}^{-S}\ ,
\end{equation}
one only integrates over the embedding variables $X^{\mu}$ (denoted here
by $X$) and the worldsheet metric $h_{\alpha\beta}$ (denoted here
by $h$).\footnote{More precisely, one has to invoke a `Faddeev--Popov
description' and introduce ghost fields and gauge-fixing terms to
evaluate the path integral.} Note that one here usually employs a
{\em Euclidean} path-integral formulation.

The path integral \eqref{pathintegral} contains in particular a
sum over all worldsheets, that is, a sum over all Riemann surfaces.
It is in this way that string interactions arise. One can then introduce
a `string coupling' constant $g$ that is related to the genus of the
worldsheet---string perturbation theory is then an expansion with
respect to $g$ (`loop expansion'). 

Demanding the absence of anomalies on the worldsheet leads to 
constraints:
\begin{itemize}
\item The background fields have to obey the `Einstein's equations'
  $+{\mathcal O}(l_{\rm s}^2)$. They can be derived from 
  an effective action, $S_{\rm eff}$, defined in the $D$-dimensional
 embedding space. 
\item The number of spacetime dimensions for the embedding space is
 restricted: One has $D=26$ for the bosonic string ($D=10$ resp. $D=11$
for the superstring).
\end{itemize}
There is an important connection between the worldsheet path integral and
the effective action: Amplitudes for scattering processes
(e.g. graviton--graviton scattering) calculated from string scattering
are identical to the field-theoretic amplitudes of the corresponding processes
which are derived from $S_{\rm eff}$ (they are 
found after a decomposition of the
form \eqref{Feynman} is made). 

Apart from the lack of experimental evidence, there are
various theoretical problems; they include
\begin{itemize}
\item There are many ways to compactify the additional spacetime
 dimensions (whose number is $D-4$). Moreover, these additionals dimensions
may be non-compact as indicated by the existence of various
`brane models'. Without a solution to this problem, no definite relation
to low-energy coupling constants and masses can be made.
\item Background independence is not yet fully implemented into string theory,
as can be recognized from the prominent role of the embedding space.
 The AdS/CFT theories discussed in recent years may come close
to background independence in some respect \cite{AGMOO}.
\item The Standard Model of particle physics, which is experimentally 
extremely well tested, has not yet been recovered from string theory.
\item What is M-theory and what is the role of the 11th dimension
which has emerged in this context?
\item Quantum cosmology has not yet been implemented into the full
theory, only at the level of the effective action (`string cosmology').
\end{itemize}

Both string theory and loop quantum gravity
exhibit aspects of non-commutative geometry. This could be relevant
for understanding space at the smallest scale. In loop quantum gravity,
the three-geometry is non-commutative in the sense that area operators
of intersecting surfaces do not commute. In string theory, for
$n$ coincident D-branes, the fields $X^{\mu}$ -- the embeddings --
and $A_a$ -- the gauge fields -- become non-commuting
$n\times n$ matrices. It has also been speculated that time could
emerge from a timeless framework if space were non-commutative
\cite{Majid}. This would be an approach very different from the
recovery of time in quantum geometrodynamics discussed above.
Quite generally, one envisages a highly non-trivial structure
of space(time) in string theory \cite{Horowitz}.

\section{Black holes}

A major application for quantum gravity is black-hole physics.
At the centre are two main questions. First, what happens at the
final evaporation phase after the breakdown of
Hawking's semiclassical approximation?
Second, how can one derive \eqref{SBH} in terms
of quantum gravitational microstates? Whereas some results have been
obtained concerning the entropy, an answer to the first question remains
elusive. It has only indirectly been attacked in the form of the
`information-loss problem'. I shall briefly discuss both issues in the
following.

\bigskip
\centerline{\bf Black-hole entropy $S_{\rm BH}$?}
\bigskip

A microscopic foundation of \eqref{SBH} has been attempted by both
loop quantum gravity and string theory. Very briefly, the situation
is as follows:
\begin{itemize}
\item {\em Loop quantum gravity}: 
The microscopic degrees of freedom are given by the spin-network states.
An appropriate counting procedure for the number of the relevant
horizon states leads to an entropy that is proportional to the
Barbero--Immirzi parameter $\beta$. The
demand for the result to be equal to
\eqref{SBH} then {\em fixes} $\beta$. Until 2004, it was believed that
the result is
\begin{equation}
\beta=\frac{\ln 2}{\pi\sqrt{3}}\quad \left(\frac{\ln 3}{\pi\sqrt{2}}\right)\ ,
\end{equation}
where the value in parentheses would refer to the choice of SO(3)
instead of SU(2). The SO(3)-value would have exhibited an interesting
connection with the quasi-normal modes for the black hole.
More recently, it was found that the original estimate for the number of
states was too small. A new calculation yields the following numerical
estimate for $\beta$ \cite{DLM}: 
 $\beta=0.237 532 \ldots$ An interpretation of this value at a more 
fundamental level has not yet been given.   
\item {\em String theory}:
The microscopic degrees of freedom are here the \emph{D-branes}, for which
one can count the quantum states in the weak-coupling limit (where 
no black hole is present). Increasing the string coupling, 
one reaches a regime where no D-branes are present, but instead one has
black holes. 
For black holes that are extremal in the relevant string-theory charges
(for extremality the total charge is equal to the mass), the number 
of states is preserved (`BPS black holes'), so the result for 
the black-hole entropy is the same as in the D-brane regime. In fact, the
result is just \eqref{SBH}, as it must be if the theory is consistent.
This remains true for non-extremal black
holes, but no result has been obtained for a generic black hole
(say, an ordinary Schwarzschild black hole). More recently, an
interesting connection has been found between the partition function,
$Z_{\rm BH}$, for a BPS black hole and a topological string amplitude,
$Z_{\rm top}$, \cite{OSV},
\begin{equation}
Z_{\rm BH}=\vert Z_{\rm top}\vert^2\ .
\end{equation}
Moreover, it has been speculated that the partition function can be
identified with (the absolute square of) the
Hartle--Hawking wave function of the universe
discussed above \cite{OVV}. This could give an interesting  
 connection between quantum cosmology and string theory.
\end{itemize}

\bigskip
\centerline{\bf Information loss for black holes?}
\bigskip

What happens during the final evaporation phase of a black hole?
This is a major question for any theory of quantum gravity to answer.
A definite answer is, however, elusive, and part of the discussion
has circled around the `information-loss problem' for black holes,
see, for example, \cite{Page,dice}.
What is the `information-loss problem'? 
According to Hawking's semiclassical calculations, a black hole
radiates with a thermal spectrum, with the temperature given by
\eqref{TBH1}. As a consequence, the black hole loses mass and shrinks.
In case that the hole completely evaporates and leaves only thermal
radiation behind, {\em any} initial state for the black hole plus
the quantum field would end with the {\em same} final state,
which would be a thermal, that is, a mixed state. This would correspond to
a maximal loss of information about the initial state. In other words:
Unitarity would be violated for a closed system, in contrast to
standard quantum theory.

Hawking originally argued that the final evaporation time,
when the black hole has reached Planck-mass size and the semiclassical
approximation breaks down, is too short for the original information
to be recovered. Therefore, so his conclusion, there must be information
loss in quantum gravity. More recently, Hawking has withdrawn his
original opinion \cite{SWH}. This switch is based on the argument
that the full evolution is unitary if also components of the 
total wave function that do {\em not} evolve
into a black hole are taken into account.

As we have seen, in full quantum gravity there is no notion of external time.
However, for an isolated system such as a black hole, one can refer
to the semiclassical time of external observers far away from the hole. 
The notion of unitarity then refers to this concept of time.
So, if the fundamental theory of quantum gravity is unitary in this sense,
there can be no information loss. Conversely, if the fundamental theory
breaks unitarity in this sense, information loss is possible. 
As long as the situation with the full theory remains open, discussions of
the information loss centre around assumptions and expectations.

However, for a black hole whose mass is much bigger than the Planck mass,
the details of quantum gravity should be less relevant. It is of
importance to emphasize that a (large) black hole is a macroscopic object.
It is therefore strongly entangled with the quantum degrees of freedom
with which it interacts. Information thereby becomes essentially non-local.
This entanglement leads to decoherence for the black hole
which thereby assumes classical properties analogously to other
macroscopic objects \cite{deco}. In particular, the black hole
itself does not evolve unitarily, cf. \cite{Zeh05}. As was shown in
\cite{DK}, the interaction of the black hole with its Hawking radiation
is sufficient to provide the black hole with classical behaviour;
strictly speaking, the very notion of a black hole emerges through
decoherence. The cases of a superposition of a black-hole state 
with its time-reversed version (a `white hole') and with a no-hole state
were considered and shown to decohere by Hawking radiation.
This does not hold for virtual black holes, which are time symmetric
and do not exhibit classical behaviour. 

In the original derivation of Hawking radiation, one starts with a
quantum field which is in its vacuum state. The thermal appearance of the
resulting Hawking radiation is then recognized from the Planckian form
of the expectation value of particle-number operator at late times.
If the state of the quantum field is evaluated on a spatial
hypersurface that enters the horizon, tracing out the degrees of freedom of
the hole interior yields a thermal density matrix
with temperature \eqref{TBH1} in the outside region.
(This is because the initial state of the field evolves into 
a two-mode squeezed state, and it is a general feature of such states
that tracing out one mode leaves a thermal density matrix for the
other mode.) However, one is not obliged to take a hypersurface that
enters the horizon. One can consider a hypersurface
that is locked at the bifurcation
point (where the surface of the collapsing star crosses the horizon). 
Then, apart from its entanglement with the black hole itself,
the field state remains pure. The observations far away from the hole
should, however, not depend on the choice of the hypersurface \cite{Zeh05}.
This is, in fact, what results \cite{CK01,dice}: The entanglement of
the squeezed state representing the Hawking radiation with other (irrelevant)
degrees of freedom leads to the thermal {\em appearance} of the field
state. It therefore seems that there is no information-loss problem
at the semiclassical level and that one can assume, 
at the present level of understanding, that the full evolution is unitary
and that `information loss' can be understood in terms of the
standard dislocalization of information due to decoherence.

A similar point of view seems to emerge from discussions invoking
string theory. In \cite{Myers} it was argued that black holes are inherently
associated with mixed states and that pure D-brane states rapidly
experience decoherence due to entanglement with other degrees of freedom.
An analogous situation (but without black holes)
was discussed in \cite{Amati}: Consider the decay of a massive string state. 
The decay spectrum of a single excited state does not show thermal
behaviour. If one, however, averages over all the degenerate states with
the same mass (justified by decoherence), the decay spectrum is Planckian.
The authors of \cite{LLM,BBJS} have argued that
black-hole geometries result for a coarse-graining over smooth
microstates. Quite generally, the thermodynamic properties of gravity
seem to be a result of decoherence.

\section{Quantum Cosmology}

\bigskip
\centerline{\bf Why quantum cosmology?}
\bigskip

Quantum cosmology is the application of quantum theory to the
universe as a whole. Independent of any particular interaction,
such a theory is needed in view of the extreme sensitivity of
quantum system to their environment, that is, to other degrees of
freedom \cite{deco}. However, since gravity is the dominating
interaction on cosmic scales, a quantum theory of gravity is needed
as a formal prerequisite for quantum cosmology.

Most work in quantum cosmology is based on the Wheeler--DeWitt
equation of quantum geometrodynamics. The method is to restrict
first the configuration space to a finite number of variables
(scale factor, inflaton field, \ldots) and then to quantize
canonically.
Since the full configuration space
of three-geometries is called `superspace', the ensuing models are
called `minisuperspace models'. Eq. \eqref{mini} is one special example
describing a Friedmann universe with a massive scalar field. 
More recently, quantum cosmology was
also discussed in the framework of loop quantum gravity.
The following issues are typically addressed within quantum cosmology:
\begin{itemize}
\item How does one have to impose boundary conditions 
in quantum cosmology?
\item Is the classical singularity being avoided?
\item How does the appearance of our classical universe emerge from
quantum cosmology?
\item Can the arrow of time be understood from quantum cosmology?
\item How does the origin of structure proceed?
\item Is there a high probability for an inflationary phase?
\item Can quantum cosmological results be justified from full quantum
gravity?
\end{itemize}
In the following I shall briefly review only few aspects. More details
can be found in \cite{OUP} as well as in the reviews
\cite{Coule,Wiltshire,Halliwell}. One particular proposal for
boundary conditions in quantum cosmology,
the Hartle--Hawking proposal, has alreay been addressed in
Sect.~3.

\bigskip
\centerline{\bf Determinism of trajectories versus determinism of waves}
\bigskip

\begin{figure}[h]
\label{fig_berlin05_6ab}
\caption{The classical and the quantum theory of gravity exhibit
drastically different notions of determinism.}
\begin{center}
\includegraphics[width=5cm]{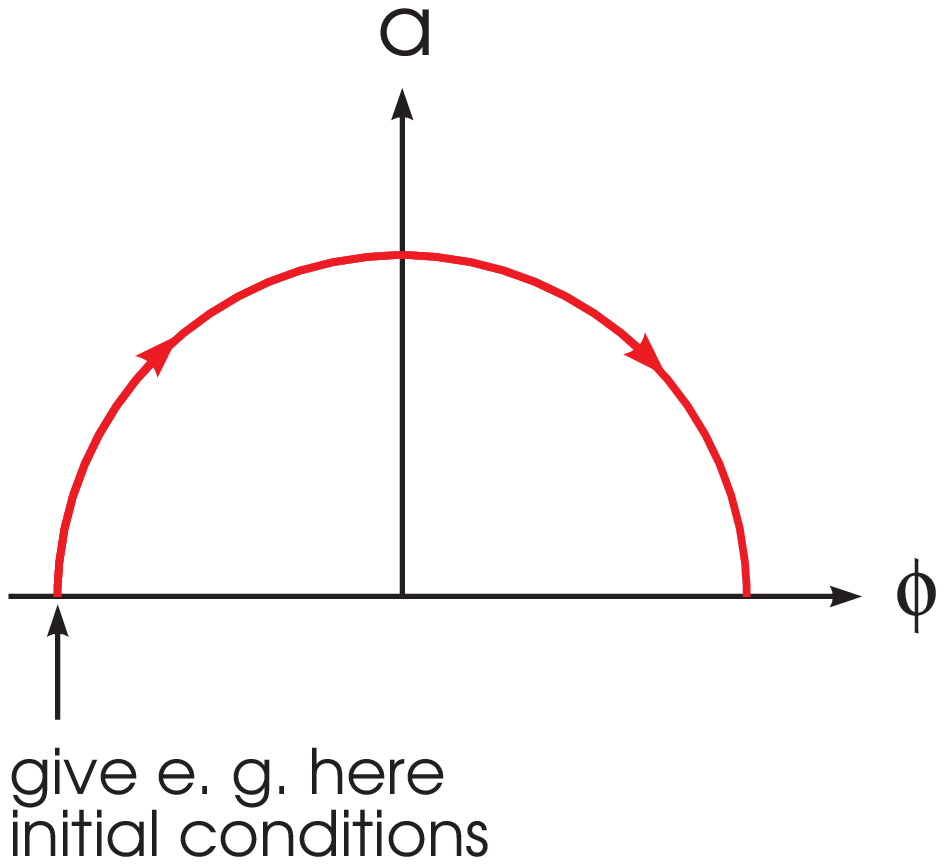}
\includegraphics[width=5cm]{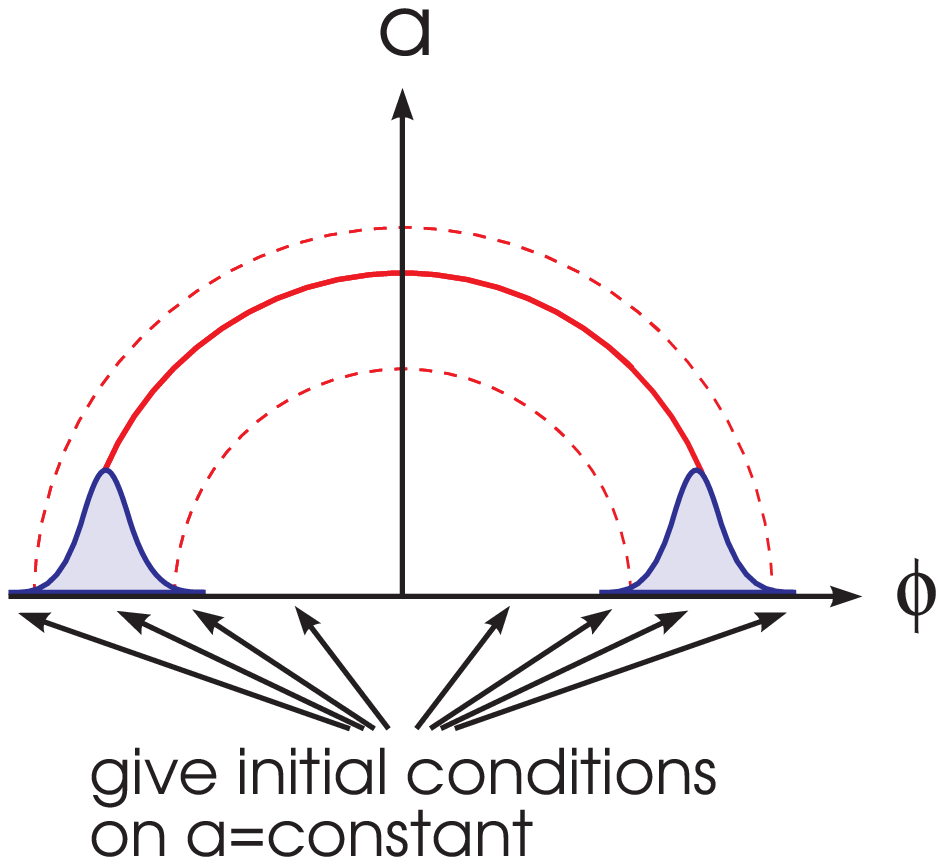}
\end{center}
\end{figure}

One interesting aspect with far-reaching consequences is the
following. The Wheeler--DeWitt equation \eqref{wdwfull} does not
contain an external time parameter. Therefore, the quantum theory
exhibits a kind of determinism drastically different from the 
classical theory \cite{OUP}. Consider a model with a two-dimensional
configuration space spanned by the scale factor, $a$, and a homogeneous
scalar field, $\phi$, see Fig.~\ref{fig_berlin05_6ab}. The classical
model be such that there are solutions where the universe expands
from an initial singularity, reaches a maximum, and recollapses to a
final singularity. Classically, one would impose $a,\dot{a},\phi,
\dot{\phi}$ (satisfying the constraint) at some $t_0$ (for example, 
at the left leg of the trajectory), and then the trajectory would be
determined. This is indicated on the left-hand side of 
Fig.~\ref{fig_berlin05_6ab}. 
In the quantum theory, on the other hand, there is no $t$.
The hyperbolic nature of \eqref{mini} suggests to impose boundary conditions
on $a=$ constant. In order to represent the classical trajectory
by narrow wave packets, the `returning part' of the packet must
be present `initially' (with respect to $a$). The determinism of the
quantum theory then proceeds from small $a$ to large $a$, not along
a classical trajectory (which does not exist). This behaviour has
consequences for the validity of the semiclassical approximation
and the arrow of time (see below).

\bigskip
\centerline{\bf Fate of classical singularity}
\bigskip

Can quantum gravity avoid the classical singularities?
So far no general criterium for singularity avoidance is available.
An important question concerns the choice of an inner product
which, for the quantum states in question, should be finite
and conserved (with respect to 
some appropriate intrinsic time or, for example
in black-hole cases, with respect to an external semiclassical time).
Some examples include
\begin{itemize}
\item No-boundary proposal: The sum is taken over regular metrics, and so
the hope is that the resulting state is singularity free.
There exist indeed minisuperspace solutions with finite inner product. 
\item Quantum dust shells can avoid the classical singularity:
A collapsing wave packet can develop into a superposition of
collapsing and expanding packets, leading to destructive interference
at the classical singularity \cite{Hajicek}. The construction
of the quantum theory is made in such a way that 
unitarity is implemented. This is not a cosmological situation,
but may serve as an appropriate analogy.
\item Loop quantum cosmology: Here the results of loop quantum gravity,
notably the discrete spectrum of the geometric operators, are imposed
on quantum cosmological models \cite{Bojowald}. The
Wheeler--De Witt equation then
becomes a \emph{difference equation} (recognizable as such near the Planck
scale, but becoming identical to the Wheeler--De Witt equation for large
scales). The inverse of the scale factor becomes 
a \emph{bounded} operator on zero volume eigenstates. One has,
for example,
\begin{equation}
\left(a^{-1}\right)_{\rm max} = \frac{32(2-\sqrt{2})}{3
  \sqrt{8\pi}l_{\rm P}}
\end{equation}
Moreover, and more important, the difference equation can be continued
through the `classical singularity'. This would be the analogue of the
unitarity condition in standard quantum theory. The results of
loop quantum cosmology have not yet been obtained from loop
quantum gravity. In \cite{BT} it is argued that issues such as
singularity avoidance must be formulated in terms of physical
quantities, that is, variables commuting with the Hamiltonian constraint.
Some criteria based on generalized coherent states are proposed there.  
\item Motivated by loop quantum cosmology, 
one can invoke new quantization methods through which cosmological
and black-hole singularities are avoided \cite{HW}.
\end{itemize}
\bigskip
\centerline{\bf Origin of irreversibility}
\bigskip

Although most fundamental laws are invariant under time reversal,
there are several classes of phenomena in Nature that exhibit an
arrow of time \cite{Zeh}. It is generally expected that there is
an underlying master arrow of time behind these phenomena, and that this
master arrow can be found in cosmology. If there is a special initial condition
of low entropy, statistical arguments can be invoked to demonstrate
that the entropy of the universe will increase with increasing size.

There are several subtle issues connected with this problem.
First, one does not yet know a general expression for the entropy
of the gravitational field, except for the black-hole entropy
\eqref{SBH}. Roger Penrose has suggested to use the Weyl tensor as
a measure of gravitational entropy, see, for example, \cite{GH}
for references and recent discussions. Second, since the very early universe
is involved, the problem has to be treated within quantum gravity.
But as we have seen, there is no external time in quantum gravity -- so
what does the notion `arrow of time' mean?

For definiteness I want to base the discussion on quantum geometrodynamics,
that is, on the Wheeler--DeWitt equation \eqref{wdwfull}.
It should be possible to do the same within loop quantum gravity
or string theory. An important observation is that the
Wheeler--DeWitt equation exhibits a
fundamental asymmetry with respect to the `intrinsic time' defined
by the sign of the kinetic term. Very schematically, one can write this
equation as
\begin{equation}
 H \, \Psi = 
\left(\frac{\partial^2}{\partial\alpha^2} + \sum_i \, \left[
-\frac{\partial^2}{\partial x_i^2}+\underbrace{V_i(\alpha,x_i)}_{\to 0\ 
{\rm for}\ \alpha
\rightarrow -\infty}\right]\right) \, \Psi = 0 \ ,
\end{equation}
where again $\alpha=\ln a$, and the $\{ x_i\}$ denote inhomogeneous
degrees of freedom describing perturbations of the Friedmann universe;
they can describe weak gravitational waves or density perturbations.
The important property of the equation is that the potential becomes small
for $\alpha\to -\infty$ (where the classical singularities would occur),
but complicated for increasing $\alpha$; 
the Wheeler--DeWitt equation thus possesses an asymmetry
with respect to `intrinsic time' $\alpha$. One can in particular impose
the simple boundary condition
\begin{equation}
\Psi \quad \stackrel{\alpha \, \to \, -\infty}{\longrightarrow}\
\prod \psi_i(x_i)\ ,
\end{equation}
which would mean that the degrees of freedom are initially {\em not}
entangled. Defining an entropy as the entanglement entropy between
relevant degrees of freedom (such as $\alpha$) and
irrelevant degrees of freedom (such as most of the $\{ x_i\}$), this
entropy vanishes initially but
increases with increasing $\alpha$ because entanglement increases
due to the presence of the potential. In the semiclassical limit where
$t$ is constructed 
from $\alpha$ (and other degrees of freedom),
cf. \eqref{semi}, entropy increases with increasing $t$. This then
\emph{defines} the direction of time and would be the origin of
the observed irreversibility in the world. The expansion of the 
universe would then be a tautology. Due to the increasing entanglement,
the universe rapidly assumes classical properties for the
relevant degrees of freedom due to decoherence \cite{deco,OUP}.
Decoherence is here calculated by integrating out the $\{ x_i\}$
in order to arrive at a reduced density matrix for $\alpha$.



This process has interesting consequences for a classically
recollapsing universe \cite{KZ,Zeh}. 
Since Big Bang and Big Crunch correspond to the same region in
configuration space ($\alpha\to-\infty$), an initial condition
for $\alpha\to-\infty$ would encompass both regions,
cf. Fig.~\ref{fig_berlin05_6ab}. This would mean
that the above initial condition would always correlate increasing
size of the universe with increasing entropy: The arrow of time would
formally reverse at the classical turning point. As it turns out, however,
a reversal cannot be observed because the universe would enter a quantum phase
\cite{KZ}. Further consequences concern black holes in such a
universe because no horizon and no singularity would ever form.

Needless to say that these considerations are speculative. 
They demonstrate, however, that interesting consequences would
result in quantum cosmology if the underlying equations were taken
seriously.

\section{Observations}

Up to now there are only expectations and hopes. Here I collect a
brief list of possible tests of quantum gravity, cf.
\cite{clausl,KM,OUP} and the references therein:
\begin{itemize}
\item Evaporation of black holes: This can only be
     observed if there are primordial black holes
      (or large extra dimensions).
\item The origin of masses and coupling constants should be understandable
    from a fundamental theory. This concerns in particular the origin of
      the cosmological constant $\Lambda$ (or dark energy).
\item Quantum gravitational corrections could perhaps be seen in
       the anisotropy spectrum of the cosmic microwave background,
 or in other cases.
\item A fundamental theory could predict
varying coupling constants, a violation of the equivalence principle
and/or a violation of Lorentz invariance.
\item A discrete microstructure 
 of space could be recognizable, for example, from
  modified dispersion relations of electromagnetic
 radiation coming from far-away objects (e.g. $\gamma$-ray bursts).
\item Experiments at the Large Hadron Collider (LHC)
could yield signatures of supersymmetry and/or higher dimensions.
\end{itemize}
But one should also keep in mind Einstein's dictum that 
only the theory itself eventually tells one what can be observed.

\section{Some questions}

I conclude with some questions:
\begin{itemize}
\item Is unification needed to understand quantum gravity?
\item Into which approaches is background independence implemented?
(In particular, is string theory background independent?)
\item In which approaches do UV divergences vanish?
\item Is there a continuum limit for path integrals?
\item Is a Hilbert-space structure needed for the full theory?
(This has an important bearing on the interpretation of quantum states.)
\item Is Einstein gravity non-perturbatively renormalizable?
(Can the cosmological constant be calculated from the IR behaviour
of renormalization-group equations?)
\item What is the role of non-commutative geometry?
\item Is there an information loss for black holes?
\item Are there decisive experimental tests?
\end{itemize}

\begin{acknowledgement}
I thank Christian Heinicke for a critical reading of this article,
and Julian Barbour, Hai Lin, and Oliver Winkler for their comments.
\end{acknowledgement}




\end{document}